\documentstyle[prl,aps,multicol,graphicx]{revtex}
\begin{document}
\newcommand\be{\begin{equation}}
\newcommand\ee{\end{equation}}

\tighten

\twocolumn[\hsize\textwidth\columnwidth\hsize\csname
@twocolumnfalse\endcsname

\title{Non-gaussian Noise in Quantum Spin Glasses
 and Interacting Two-level Systems} 
\author{A. K. Nguyen$^{1}$ and S. M. Girvin$^{1,2}$} 
\address{{}$^1$Department of Physics, Indiana University, Bloomington, 
IN 47405, USA}
\address{{}$^2$ Institute of Theoretical Physics, University of California, 
Santa Barbara, CA 93106, USA}

\draft
\maketitle

\begin{abstract}
%
%
We study a general model for non-gaussian $1/f$ noise based on an
infinite range quantum Ising spin system in the paramagnetic state, or
equivalently, interacting two-level classical fluctuators.  We
identify a dilatation interaction term in the dynamics which survives
the thermodynamic limit and circumvents the central limit theorem to
produce non-gaussian noise even when the equilibrium distribution is
that of {\em non-interacting} spins. The resulting second spectrum
(`noise of the noise') itself has a universal $1/f$ form which we
analyze within a dynamical mean field approximation.
\end{abstract}
\vskip2pc]


Noise measurements have   
long been an important probe of the still poorly understood
dynamics of spin glasses \cite{Weissman:RMP93}, structural glasses and
two-level systems \cite{Walther:PRB98,aSi}.  
Interest in $1/f$ noise
\cite{Weissman:RMP88,Dutta:RMP81} has been further enhanced with
recent progress in the development of nano-scale electronic devices
such as the single electron transistor \cite{Schoelkopf:Science98} and
precision charge pumps \cite{Kautz:PRB00} which are extremely
sensitive to offset charge fluctuations \cite{Zimmerman:PRB97}.  Such
noise also plays a significant role in determining the coherence time
of existing ion-trap realizations of quantum bits
\cite{Turchette:PRA00} and is likely to be even more important in
other proposed solid state realizations of qubits such as quantum dots
\cite{Loss:PRA98}, Cooper pair boxes \cite{Nakamura:Nature99}, and
Josephson junctions \cite{Makhlin:RMP01}.  A constant loss tangent in
a capacitor leads to a $1/f$ noise spectrum \cite{Israeloff:PRB96} and
loss tangents even as a small as $10^{-5}$ would have serious
consequences in charge-based quantum bits.  Ubiquitous non-equilibrium
noise may also be implicated in the coherence time of electrons in
wires and films observed via weak localization and via mesoscopic
persistent currents \cite{webb}.


We focus here on non-gaussian fluctuations which 
are usually connected with critical phenomena
\cite{Goldenfeld:Book92} or systems having only a
few degrees of freedom \cite{Weissman:RMP88}.  
We develop a simple model with a {\em large} number of
degrees of freedom and identify interaction terms that circumvent the central
limit theorem and produce non-gaussian fluctuations in the
thermodynamic limit even in the paramagnetic state far above
any critical point.  We analyze the non-gaussian fluctuations using the
second spectrum 
(`noise of the noise') \cite{Restle:PRB85,Weissman:RMP93,Seidler:PRB96}.
In addition to being of interest for the study of two-level systems, the
dynamical model we present is also relevant to models of 
quantum spin glass behavior
\cite{quantumspinglass}.

Consider $N$ interacting two-level fluctuators (TLFs) labeled by
Ising spin variables $S_i = \pm 1$ whose flipping rates obey
\begin{eqnarray}
  \frac{1}{\tau_i(S_i)} = e^{ - \big [ \tilde{D}_i + h H_i S_i +
  \frac{1}{\sqrt{N}} \sum_{j \ne i} \gamma \Gamma_{ij}S_j S_i \big ]}.
\label{eq:corner}
\end{eqnarray}
This classical dynamics is equivalent 
to the imaginary time dynamics of a quantum spin Hamiltonian
\begin{equation}
{\bf \mathcal{H}} = - \sum_{i=1}^N \! \mbox{\boldmath $\sigma$}_i^x 
\frac{1}{\tau_i(\sigma^z_i)}
\label{NG.Ham}
\end{equation}
with the Ising variables $S$ replaced by Pauli spin matrices
$\sigma^z$.  For a spin glass, $h H_i$ would typically represent a
uniform external magnetic field, i.e. $H_i = 1$.  For a structural
glass the $h H_i$ represent random local site energy differences for
the TLFs.  $\gamma \Gamma_{ij}$ is the spin-spin interaction energy
and $\tilde{D}_i = \Delta_i + \frac{1}{\sqrt{N}} \sum_{j\ne i}
[\lambda\Lambda_j + gG_{ij} ]S_j $ is the barrier height which
fluctuates around its bare value $\Delta_i$ due to coupling between
the TLFs.  We take $H_i$, $\Lambda_i$, $G_{ij}=G_{ji}$ and
$\Gamma_{ij}=\Gamma_{ji}$ to be random and uniformly distributed on
$[-1,1]$, as in the infinite-range
Sherrington-Kirkpatrick model \cite{quantumspinglass}. 

In steady state, detailed balance gives us the single-spin probability
ratio
\begin{eqnarray}
  \frac{P(S_i = 1)}{P(S_i = -1)} = e^{2 h H_i + 2
  \frac{\gamma}{\sqrt{N}} \sum_{j\ne i} \Gamma_{ij} S_j}.
\end{eqnarray}
Note that this is completely independent of $\tilde{D}_i$ which
controls various features of the dynamics but has no effect on the
equilibrium statistical mechanics.

If and only if $\Gamma$ is symmetric, we can write the full
equilibrium probability distribution in terms of a Boltzmann factor
derived from a Hamiltonian
\begin{eqnarray}
  P[S] = \frac{1}{Z} e^{h \sum_i H_iS_i + \frac{\gamma}{\sqrt{N}}
  \sum_{i < j} \Gamma_{ij} S_i S_j},
\end{eqnarray}
where $Z$ is the partition function.  Notice that the matrix $G$ need
{\em not} be symmetric.  The approach to equilibrium might depend on
this, but the equilibrium does {\em not}.

For large enough $\gamma$ (low enough temperature) the system
undergoes a glass transition into a frozen state.  Near this
transition the dynamics may also become non-gaussian, but we restrict
our attention here to the high temperature paramagnetic regime where
$\gamma$ is small.  {\em Our central result is that even if $\gamma=0$
so that the equilibrium distribution is that of { non-interacting}
spins, the second spectrum indicates non-gaussian dynamics for any
$\lambda \ne 0$.}

The term $\sum_{j \not = i} \lambda \Lambda_j S_j$ is a `dilatation'
term. It causes all the barriers to collectively and simultaneously
increase or decrease in a correlated manner. In contrast to this,
interactions through the $g G_{ij}$ and $\gamma \Gamma_{ij}$ terms
cause barrier fluctuations which are statistically independent on
different sites (at least for small $g$ and $\gamma$ so that the
system is above the critical point). As a consequence of the central
limit theorem, one expects that only the dilatation term survives 
to produce non-gaussian fluctuations in the thermodynamic limit
of this infinite-range model. 

A possible physical realization of such a dilatation term is the
elastic interaction between TLFs. In addition to the dipole term one
could have an expansion of the lattice coupled to the impurity
positions \cite{Sethna:pc}. The motion of a defect then changes the
hydrostatic pressure by an amount proportional to $\frac{1}{\mathcal
V}$ throughout the volume ${\mathcal V}$ \cite{Landau:Book59}. Since
the barriers would tend to go up uniformly with increasing hydrostatic
pressure, a long-range dilatation term with 
an amplitude $\sim {N^{-1}}$ would be
present \cite{Sethna:pc}.  Such a term thus could be important in small
grains.  We have chosen for simplicity
to scale the interaction with $N^{-\frac{1}{2}}$
in order to obtain a sensible thermodynamic limit in the present model.

To ensure that the power spectrum of the model decays as $1/f$, we
take the bare barriers to have a flat distribution on a fixed interval
\cite{Weissman:RMP88} $\Delta_i \!\in\! [\Delta^{\rm min}, \Delta^{\rm
max}]$. We carry out simulations using Eq.~(\ref{eq:corner}). To
average, each ``measurement'' was repeated $1600$ times, where
$\Delta_i$, $\Lambda_i$, $G_{ij}$ and $\Gamma_{ij}$ are assigned new
values each time. Using $\delta t$ as a discretizing time interval,
the noise signal is defined as $ V(t~\delta t) = N^{-1/2} \!
\sum_{i=1}^{N} \!  S_i(t~\delta t)$. Henceforth, we set $\delta t =
1$.
\begin{figure}[h]
  \begin{picture}(0,185)(0,0)
      \put(-95,-355){\includegraphics[angle=0,scale=0.7] {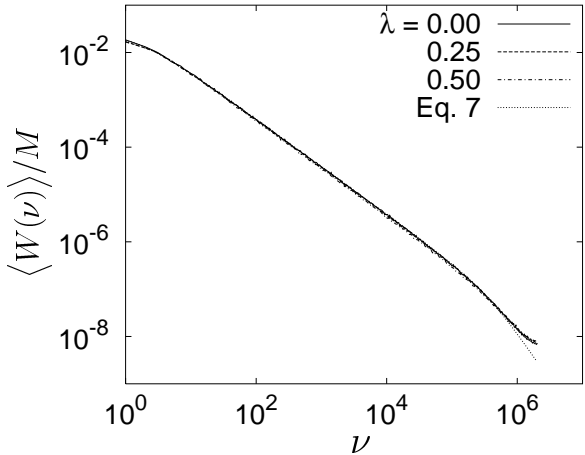}}
  \end{picture}
  \caption{Power spectrum as a function of frequency for a
  non-interacting ($\lambda = g = h = \gamma = 0$) and two interacting
  ($\lambda = g = h = \gamma = 0.25, 0.5$) cases. Other parameters
  are: $\Delta^{\rm max} = 14$, $\Delta^{\rm min} = 1$, $M = 2^{22}$
  and $N = 100$. $\big \langle W(\nu) \big \rangle$ is found to be
  independent of $N$. Thin dotted line is a plot of the analytical
  expression for the non-interacting case,
  Eq.~(\ref{Pow.Spec.Res}). Typical barrier fluctuations for the
  ($\lambda = g = h = \gamma = 0.5$) case are $\sim\!\! 10 \%$ and
  $\sim\!\!  100 \%$ of $\Delta^{\rm max}$ and $\Delta^{\rm min}$,
  respectively, indicating weak or moderate interactions.}
\label{Pow.Spec.Fig}
\end{figure}
%
In this letter, $t, t_2$ and $\nu, \nu_2$ are dimensionless integer
times and frequencies, respectively. The normalization $N^{-1/2}$
ensures that $\big \langle V^2(t) \big \rangle = 1$, independent of
$N$.

To study the time-time correlation of the noise, we use the power
spectrum defined as $\big \langle W(\nu) \big \rangle = \big
\langle \big | M^{-1/2} \sum_{t=1}^M e^{2 \pi i \nu t/M} V(t) \big |^2
\big \rangle$, where $M$ is the length of the time sequence. We see in
Fig.~\ref{Pow.Spec.Fig} that the {\em power spectrum is insensitive to
weak interactions}.

\begin{figure}[h]
  \begin{picture}(0,185)(0,0)
      \put(-90,-355){\includegraphics[angle=0,scale=0.7] {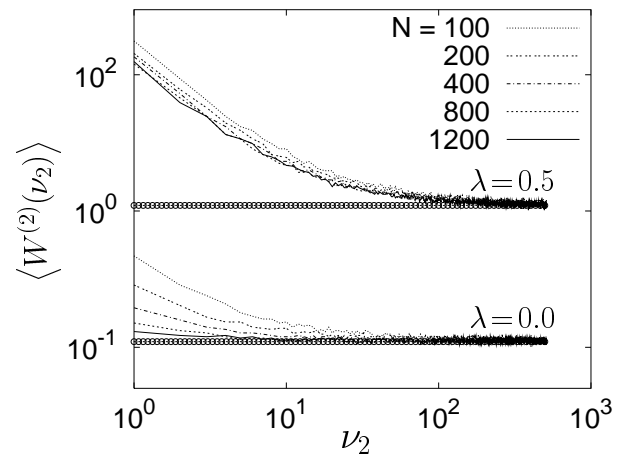}}
  \end{picture}
  \caption{Second spectrum as a function of frequency for $\lambda =
  0$ and $0.5$ with $N = 100,200,400,800,1200$. Other parameters are:
  $g = h = \gamma = 0.5$, $M_1 = M_2 = 2^{10}$, $\Delta^{\rm max} =
  15$, $\Delta^{\rm min} = 0.5$, $\nu^{\rm min} = 125$ and $\nu^{\rm
  max} = 175$. Data for the $\lambda = 0$ case is shifted down one
  decade for clarity. Dots ($\odot$) show the second spectrum for the
  non-interacting (gaussian) case.}
\label{Sec.Spec.Fig}
\end{figure}

To study non-gaussian properties of the noise, we use the second
spectrum introduced by Restle et
al.~\cite{Restle:PRB85,Weissman:RMP93,Seidler:PRB96}. Given a time
sequence of length $M$, we split it into $M_2$ shorter time sequences
of length $M_1$ each, and define the second spectrum as
\begin{eqnarray}
   \Big \langle W^{(2)}(\nu_2) \Big \rangle &=& \left \langle \bigg
   |\frac{1}{\sqrt{M_2}} \sum_{t_2=1}^{M_2} e^{2\pi i \nu_2 t_2/M_2}
   \!\!\! \sum_{\nu = \nu^{\rm min}}^{\nu^{\rm max}} \!\! W_{t_2}(\nu) \bigg
   |^2 \right \rangle, \!\!\!\!\!\!
\nonumber \\ 
   W_{t_2}(\nu) &=& \Bigg | \frac{1}{\sqrt{M_1}} \sum_{t=t_2
   M_1}^{(t_2+1) M_1} e^{2\pi i \nu t/M_1} V(t) \Bigg |^2.
\label{SecSpec} 
\end{eqnarray}
The quantity $W_{t_2}(\nu)$ is a single shot ``power spectrum'' of
$V(t)$ in the time window $t_2$. Physically, the second spectrum is a
spectrum of the power spectrum's wandering within a chosen frequency
band, $[\nu^{\rm min},\nu^{\rm max}]$. It is a four-point correlation
function which probes the noise of the noise
\cite{Restle:PRB85,Weissman:RMP93,Seidler:PRB96}. For the
non-interacting (gaussian) case the second spectrum becomes flat
\cite{Weissman:RMP88} and takes the value $\sum_{\nu=\nu^{\rm
min}}^{\nu^{\rm max}} \big \langle W(\nu) \big \rangle^2$.  We see in
Fig.~\ref{Sec.Spec.Fig} that for $\lambda = 0$ the second spectrum
decreases for increasing $N$, merging into its non-interacting
(gaussian) counter part for large $N$. However, for $\lambda = 0.5$
the second spectrum is independent of $N$ for large $N$.
Thus, {\em the dilatation term in Eq.~(\ref{eq:corner}) is the
only term that survives the central limit theorem}.
All the other terms produce only gaussian fluctuations in the
thermodynamic limit.

We now calculate the power spectrum and second spectrum analytically.
First, we set $g = h = \gamma = 0$ since, as shown above,
 these terms do
not contribute to non-gaussian properties of the model and the power
spectrum is insensitive to weak interactions.  Second, we set
$\Delta_i = (\Delta^{\rm max}-\Delta^{\rm min})i/N + \Delta^{\rm min}$. Third,
we replace $\Lambda_i \!\in\! [-1,1]$ by $\Lambda_i =
1$. Mathematically, it is always possible to make a variable
(`gauge') transformation $S_j' = {\rm sign}(\Lambda_j) S_j$ such that all
$\Lambda_i$
are positive in the new basis.  Simulations have shown that the
results for these two cases are qualitatively the same, with the
second spectrum of the $\Lambda_i = 1$ case somewhat larger as
expected.


We now use the Suzuki-Trotter decomposition
\cite{Suzuki:PTP76} and recast the quantum model,
Eq.~(\ref{NG.Ham}), into a classical model with the partition function
\begin{eqnarray}
   & & Z = \left \{ \prod_{i=1}^N \prod_{t=1}^M \sum_{S_i(t)}
       \sum_{\tilde{V}(t)} \delta \Big [\tilde{V}(t) \!-\!\!
       \frac{1}{\sqrt{N}} \sum_j S_j(t) \Big ] \right \} \times
\label{Part.Func} \\
   & & \exp \left [ \sum_{i,t} \bigg \{ K_i(t) S_i(t) S_i(t\!+\!1) +
       \frac{1}{2} \ln \Big [ \frac{1}{2} \sinh [2 \epsilon_i(t) ]
       \Big ] \bigg \} \right ].  \nonumber
\end{eqnarray}
Here, $\tilde{V}(t)$ is an auxiliary field used to decouple the
spins. Furthermore, $\epsilon_i(t) = \frac{\beta}{M} e^{-\Delta_i -
\lambda \tilde{V}(t)}$ and $K_i(t) = \frac{1}{2} \ln \big ( \coth [
\epsilon_i(t)] \big )$, where $\beta = 1/k_B T$ is the inverse
temperature. In Eq.~(\ref{Part.Func}), we have replaced $\delta \big
[\tilde{V}(t) \!-\!\!  N^{-1/2} \! \sum_{j \not = i} S_j(t) \big ]$ by
$\delta \big [\tilde{V}(t) \!-\!\!  N^{-1/2} \! \sum_{j = 1}^N S_j(t)
\big ]$ and, thereby, introduced an error of order $O(N^{-1/2})$ which
vanishes in the $N \rightarrow \infty$ limit.

Using Eq.~(\ref{Part.Func}), the power spectrum for the non-interacting
case may in the large $N$ and large $M$ limit be shown to take the
well known form
\begin{eqnarray}
   \Big \langle W(\nu;\lambda = 0) \Big \rangle = \frac{
   \mbox{atan}( \frac{e^{-\Delta^{\rm min}}}{\pi \nu/M} \frac{\beta}{M}) -
   \mbox{atan}( \frac{e^{-\Delta^{\rm max}}}{\pi \nu/M} \frac{\beta}{M})
   }{(\Delta^{\rm max}-\Delta^{\rm min}) (\pi \nu/ M)}.
\label{Pow.Spec.Res}
\end{eqnarray}
From Eq.~(\ref{Pow.Spec.Res}), we see that for intermediate frequencies
$\big \langle W(\nu;\lambda = 0) \big \rangle \sim 1/\nu$, independent of
the temperature. Henceforth, we set $\beta/M = 1$ so that the
analytical and simulation results are directly comparable.

We plot Eq.~(\ref{Pow.Spec.Res}) in Fig.~\ref{Pow.Spec.Fig} using the
same parameters as in the simulations. There is nice agreement between
the simulation and analytical results. At the highest frequencies,
$\nu > 10^6$, results from the simulations are higher compared to the
analytical result. This is expected since in the simulations the time
sequences are discrete, while in the analytical calculations a
continuum description is used to integrate out the sums. It is
precisely at these highest frequencies that the difference between the
two methods becomes important.

We now apply the dynamical mean-field approximation (DMFA)
\cite{quantumspinglass} by
replacing the formal averaging over $\tilde{V}$ in Eq.~(\ref{Part.Func})
with the requirement that all 
(multi-point)   
correlation functions of
$\tilde{V}(t)$ are self-consistently
identical to the corresponding correlations
functions of $V(t)$. For example, $\big \langle \tilde{V}(t)
\tilde{V}(t') \big \rangle_{\tilde{V}} = \big \langle V(t) V(t')
\big \rangle_S$.  Here, $\big \langle \big \rangle_S$ and $\big
\langle \big \rangle_{\tilde{V}}$ are averages over the spin
configurations and the auxiliary field, respectively.  As for other
infinite-range spin models
\cite{quantumspinglass}, 
this approximation becomes exact in the $N \rightarrow
\infty$ limit.

For a classical 
1D 
Ising model $\big \langle S(t) S(t') S(t'') S(t''')
\big \rangle = \big \langle S(t) S(t') \big \rangle \, \big \langle
S(t'') S(t''') \big \rangle$, given that the overlap between the time
intervals $[t,t']$ and $[t'',t''']$ is zero. Thus, the term $\big
\langle W_{t_2}(\nu) ~ W_{t_2'}(\nu') \big \rangle_S$ in
Eq.~(\ref{SecSpec}) may be replaced by $\big \langle W_{t_2}(\nu) \big
\rangle_S ~ \big \langle W_{t_2'}(\nu') \big \rangle_S +
O(M_2^{-1/2})$, where the error $O(M_2^{-1/2})$ vanishes in the $M_2
\rightarrow \infty$ limit. Using Eq.~(\ref{Part.Func}) and the DMFA we obtain
\begin{eqnarray}
   \Big \langle W_{t_2}(\nu;\lambda) \Big \rangle_S = \frac{1}{M_1}
   \sum_{t',t''=t_2 M_1}^{(t_2+1)M_1} e^{2 \pi i \nu (t'-t'')/M_1}
   \times \nonumber \\ \frac{1}{N} \sum_{i=1}^N \exp \left [-2
   e^{-\Delta_i} \sum_{t=t'}^{t''} e^{-\lambda \tilde{V}(t)} \right ].
\label{Pow.Spec.m.def}
\end{eqnarray}
We split $\tilde{V}$ into fast and slow components, $\tilde{V} =
\tilde{V}_{f} + \tilde{V}_s$, and then assume that $\tilde{V}_s$ is
constant over the time interval $[t',t'']$ whereas $\sum_{t=t'}^{t''}
e^{-\lambda \tilde{V}_f(t)}$ is replaced by its second cumulant
average.  The second cumulant depends only on the first spectrum,
which from Fig.~(\ref{Pow.Spec.Fig}) we know is universal and
independent of the all couplings including $\lambda$.
In Eq.~(\ref{Pow.Spec.m.def}) we may therefore make the approximation
$\sum_{t=t'}^{t''} e^{-\lambda \tilde{V}(t)} \simeq |t' - t''|
e^{-\lambda \tilde{V}_s(t_2) + \frac{\lambda^2}{2}\xi}$ where $\xi$ is
an unimportant constant which will not contribute to the second
spectrum computed to order $\lambda^2$ below.
Applying the approximation to
Eq.~(\ref{Pow.Spec.m.def}) we find
\begin{eqnarray}
   \Big \langle W_{t_2}(\nu;\lambda) - W(\nu;0) \Big \rangle_S \simeq
   \frac{e^{\Delta^{\rm min}}(1-e^{\lambda \tilde{V}_s(t_2) -
   \frac{\lambda^2}{2}\xi})} {(\Delta^{\rm max} \! -\Delta^{\rm
   min})},
\label{SP.shift}
\end{eqnarray}
within the frequency band $1 \ll \nu \ll \frac{M_1}{\pi}
e^{-\Delta^{\rm min} - \lambda}$. Here, $\nu \in
[0,\frac{M_1}{2}]$.  Eq.~(\ref{SP.shift}) shows that
the shape of $\big \langle W_{t_2}(\nu;\lambda) \big \rangle_S$ is not
modified by the interactions between the fast and the slow
fluctuators. However, its amplitude shifts up or down depending on the
configuration of the slow TLFs during the short time interval
$t_2$. In other words, {\em the wandering of the power spectrum is
controlled by the signal of the slow fluctuators}.  Because the barrier
distribution is flat, there is no net effect on the first spectrum, but the
mean square wandering produces the second spectrum.

\begin{figure}[h]
  \begin{picture}(0,185)(0,0)
      \put(-95,-355){\includegraphics[angle=0,scale=0.7] {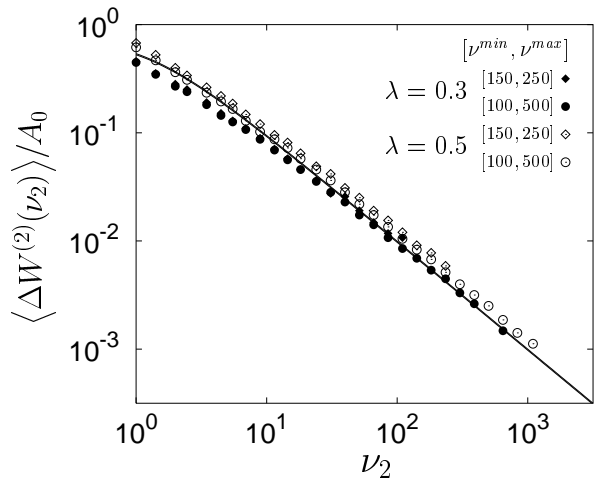}}
  \end{picture}
  \caption{Second spectrum minus it gaussian counterpart as a function
  of frequency for $g=h=\gamma=0$, $\Delta^{\rm max} = 17$,
  $\Delta^{\rm min} = 1$, $M_1 = 2^{14}$, $M_2 = 2^{12}$ and
  $N=100$. $\big \langle \Delta W^{(2)}(\nu_2) \big \rangle$ is found
  to be independent of $N$ for $N \ge 100$ when $g=h=\gamma=0$. Solid
  line is a plot of the analytical expression
  Eq.~(\ref{Sec.Pow.Res}).}
\label{Sec.Pow.Fig}
\end{figure}

Expanding Eq.~(\ref{SP.shift}) to second order in $\lambda$ and
ensemble averaging over the fluctuations of $\tilde V_s(t_2)$ we see
that the non-gaussian contribution to the second spectrum is, up to a
simple normalization factor, determined by the slow part of the first
spectrum:
\begin{eqnarray}
  & & \Big \langle \Delta W^{(2)}(\nu_2;\lambda) \Big
  \rangle_{S,\tilde{V}_s} \equiv ~ \Big \langle W^{(2)}(\nu_2;\lambda)
  - W^{(2)}(\nu_2;0) \Big \rangle_{S,\tilde{V}_s}
\nonumber \\ 
  & & \simeq \left[\lambda e^{\Delta^{\rm min}} \frac{(\nu^{\rm
  max}-\nu^{\rm min})} {(\Delta^{\rm max}-\Delta^{\rm min})}\right]^2
  W(\nu_2;\lambda=0) \simeq \frac{A_0}{\nu_2},
\label{Sec.Pow.Res}
\end{eqnarray}
where $A_0 = \frac{[\lambda e^{\Delta^{\rm min}}(\nu^{\rm
max}-\nu^{\rm min})]^2}{[\Delta^{\rm max} - \Delta^{\rm min}]^3
[2/M_2]}$. From Eq.~(\ref{Sec.Pow.Res}) and the universality of the
first spectrum shown in Fig.~(\ref{Pow.Spec.Fig}) we see that, apart
from the prefactor, the shape of the non-gaussian part of second
spectrum is independent of all interaction parameters as well as the
size and position of the frequency window used to probe it, provided
we restrict the frequency band to be inside $1 \ll \nu \ll
\frac{M_1}{\pi} e^{- \Delta^{\rm min} -\lambda}$. In
Fig.~\ref{Sec.Pow.Fig} we plot simulation results for four 
different sets of parameters. We see that the simulation results
collapse onto the line predicted by the analytical result.
The small deviations may be explained by the fact that $\lambda$ is
not infinitesimal, by practical limitations on the available dynamical
range which cause the inequality below Eq.~(\ref{SP.shift}) to be
imperfectly satisfied, and by the unavoidable poor sampling statistics
for the slowest fluctuators.

In the present model, the second spectrum has been found to decay
close to $1/f$; that is, the $1/f$ noise itself has $1/f$ noise.  This
is the same result seen experimentally in
semiconductors \cite{Voss:PRB76,Weissman:RMP88,aSi},
discontinuous metal films \cite{Voss:PRB76},
spin glasses \cite{Weissman:RMP93} and carbon resistors
\cite{Seidler:PRB96}. Our model is an infinite-range interaction model
in which we have chosen the simplest possible dynamics that
circumvents the central limit theorem to yield non-gaussian noise even
in the thermodynamic limit of the weakly correlated paramagnetic
phase. To the extent that the Ising variables we use could represent
entire droplets of spins, our model could also potentially be relevant
to the 
\noindent
strongly correlated regime of spin glasses
\cite{Fisher:PRB88b}.

We thank M. Weissman, J. Sethna, M. Kardar, J. Carini and N. Ma, for
useful discussions. This work is supported by the Norwegian Research
Council 13527/410 and by NSF DMR-0087133 and the ITP at UCSB under NSF
PHY-9907949.

\end{document}